\documentclass[pre,twocolumn,superscriptaddress,showpacs]{revtex4}

\usepackage{amsmath}
\usepackage{amsfonts}
\usepackage{amssymb}
\usepackage{graphicx}
\newcommand{\be}{\begin{equation}}
\newcommand{\ee}{  \end{equation}}
\newcommand{\ba}{\begin{eqnarray}}
\newcommand{\ea}{  \end{eqnarray}}

\begin{document}
\title{Decay of the Loschmidt echo in a time-dependent environment}
\author{F. M. Cucchietti}
\affiliation{T-4, Theory Division, MS B213, Los Alamos National Laboratory, Los Alamos, New
Mexico 87545, USA}
\author{C. H. Lewenkopf}
\affiliation{Instituto de F\'{\i}sica, Universidade do Estado do Rio de Janeiro, 20559-900
Rio de Janeiro, Brazil}
\author{H. M. Pastawski}
\affiliation{Facultad de Matem\'{a}tica, Astronom\'{\i}a y F\'{\i}sica, Universidad
Nacional de C\'{o}rdoba, Ciudad Universitaria, 5000 C\'{o}rdoba, Argentina}

\begin{abstract}
We study the decay rate of the Loschmidt echo or fidelity in a chaotic system under a 
time-dependent perturbation $V(q,t)$ with typical strength $\hbar/\tau_{V}$. 
The perturbation represents the action of an uncontrolled environment interacting 
with the system, and is characterized by a correlation length $\xi_0$ and a
correlation time $\tau_0$. 
For small perturbation strengths or rapid fluctuating perturbations, the Loschmidt echo 
decays exponentially with a rate predicted by the Fermi 
Golden Rule, $1/\tilde{\tau}=  \tau_{c}/\tau_{V}^2$, where typically 
$\tau_{c} \sim \min[\tau_{0},\xi_0/v]$ with $v$ 
the particle velocity. Whenever the rate $1/\tilde{\tau}$ is larger than the Lyapunov 
exponent of the system, a perturbation independent Lyapunov decay regime arises.
We also find that by speeding up the fluctuations (while keeping the perturbation strength 
fixed) the fidelity decay becomes slower, and hence, one can protect the system against 
decoherence.
\end{abstract}
\date{\today}

\pacs{03.65.Yz, 05.45.Mt}
\maketitle

\section{Introduction}
\label{sec:introduction}

The time-evolution of a quantum system is quite robust to changes of the system
initial conditions, irrespective of the nature of the underlying dynamics \cite{Izrailev}.
This is in deep contrast to classical evolution, particularly that of a chaotic system.
In a seminal paper, Peres \cite{Peres84} noticed that quantum time-evolution
can be sensitive to the differences between chaotic and integrable dynamics
in a peculiar set up: One needs to examine the overlap of identically prepared states, 
but evolved with slightly different Hamiltonians. This overlap, called Loschmidt echo (LE) 
or fidelity  \cite{Jalabert01}, measures the recovery obtained when a wave packet evolves 
for a time $t$, followed by a backwards evolution with a perturbed Hamiltonian for 
the same time interval. 

A considerable number of investigations has been devoted to study the 
interesting and intricate phenomena related to the LE, in particular the different
regimes that arise depending on the perturbation strength.
For very small perturbations, the LE is 
described by standard perturbation theory and a Gaussian decay is observed. 
For stronger perturbations, where perturbation theory breaks down, large phase 
fluctuations \cite{Jalabert01} lead to an exponential decay of the LE 
described by the Fermi Golden Rule (FGR) \cite{Jacquod01,CookSmooth}. 
For even stronger perturbations, but still weak in the classical sense, a semiclassical 
analysis yields an exponential LE decay that does not depend on the perturbation strength: 
The decay rate is determined by the Lyapunov exponent that characterizes the classical 
counterpart of the unperturbed system \cite{Jalabert01}. The latter two cases are called 
the FGR and Lyapunov regimes respectively, where 
the LE decay rate is the minimum between the width of the local density of 
states (LDOS), as given by the FGR, and the Lyapunov exponent 
\cite{Jalabert01,Jacquod01,CookSmooth}. 
These predictions were verified numerically in a number of systems
\cite{Jacquod01,CookSmooth,CookLorentz,Wisniacki02,Benenti02}.
The theory is successful to the extend that, by analyzing the LE 
decay, the quantum evolution of a system can be used to quantitatively assess
its classical Lyapunov exponent \cite{CookSmooth,Emerson02}.

The theory was later extended to classically integrable systems 
\cite{Jacquod02}, in which case a power law like decay is predicted.
This result is still somewhat controversial \cite{ProsenInt}, since
as a rule integrable systems display non-generic features 
\cite{weinstein-integ}. 
In any event, these works indicate that the LE decay is very different 
whether the underlying classical system has a chaotic, integrable, or  
even mixed phase space \cite{Weinstein02}. 

Albeit this wealth of interesting results, so far the theory of the LE non-perturbative 
regime has only dealt with time independent perturbations. 
The most probable motivation for this restrictive choice can be traced back to the experiments 
that triggered the research on the LE problem \cite{spinReversal98, spin-Losch00}: 
They studied the time reversal of many-spin dynamics, where the perturbation is simply 
a static part of the Hamiltonian.  

Numerous physical situations call for an extension of the LE theory that  
accounts for a time-dependent perturbation.
Let us explicit mention a few. 
Experimentally, a subsystem selected from a large spin system with many-body interactions 
can be represented as immersed in an external fluctuating potential \cite{QZE98} -- 
the same approximation holds whenever the uncontrolled degrees of freedom are those of 
an environment with complex dynamics. Formally, the current analytical description 
contrasts with numerical results \cite{Schomerus} observed in periodically 
kicked one-dimensional models \cite{Jacquod01, Benenti02}, where the perturbation can be 
interpreted as time dependent. 
The need is further stressed by the relevance of the LE to quantum computation 
\cite{Berman02, BenentiQC,Pascazio}, decoherence in open systems 
\cite{Zurek-Cook, Zurek-RMP,Gorin04}, and mesoscopic physics \cite{Stern}.
Indeed, the decay of the LE is related to the decay of quantum correlations and the 
quantum-classical correspondence, as can be shown using the Wigner function representation 
\cite{Ozorio02, CookLorentz03, Zurek-Cook, Prozen05}.

In this work we use the semiclassical approximation to derive the LE 
decay in the presence of a time-dependent perturbation, generalizing 
the approach presented in Ref.~\onlinecite{Jalabert01}. 
We show that the existence of a LE perturbation-independent regime 
is quite generic. For that purpose, instead of using a particular model, 
we use a statistical approach. 
We obtain a closed expression for the LE decay in the FGR regime using
simple assumptions on the perturbation autocorrelation function. 
We conclude by discussing the different limits of our results and the seemingly 
strange feature that faster fluctuations of the perturbation or stronger 
chaos in the system lead to a slower decay of the Loschmidt echo

\section{Loschmidt echo in a time-dependent environment}

The object of interest, the Loschmidt echo, is defined as:
\begin{equation}
M(t)=\left\vert \langle\psi_{0}|U(t_{0},t)U_{0}(t,t_{0})|\psi_{0}
\rangle\right\vert ^{2}, \label{eq:defM(t)}
\end{equation}
where $|\psi_{0}\rangle$ is an arbitrary wave packet prepared at time $t_{0}$.
For simplicity, and in line with
Ref.~\onlinecite{Jalabert01}, we choose the initial state $\left\vert 
\psi_{0}\right\rangle$ as a Gaussian wave-packet centered at an arbitrary point
$\mathbf{r}_{0}$ with dispersion $\sigma$ and initial momentum $\mathbf{p}_{0}$. 
This restricted choice can be relaxed by considering other kinds of localized 
states in phase space \cite{Casati02,Vanicek03}, evolved states \cite{Jacquod02}, 
and even eigenstates of $H_{0}$ \cite{Jacquod01,Ozorio02}.
In Eq.~(\ref{eq:defM(t)}), $U_{0}$ is the standard time evolution operator, namely
\begin{equation}
U_{0}(t,t_{0})=T\exp\left(  -\frac{\mathrm{i}}{\hbar}\int_{t_{0}}
^{t}\mathrm{d}t^{\prime}H_{0}(t^{\prime})\right),
\end{equation}
where $T$ is the time ordering operator, while
\begin{equation}
U(t_{0},t)=\overline{T}\exp\left(  -\frac{\mathrm{i}}{\hbar}\int_{t}^{t_{0}
}dt^{\prime}H(t^{\prime})\right),
\end{equation}
with $\overline{T}$ the inverse time ordering operator. 
Equation (\ref{eq:defM(t)}) is also viewed as the fidelity of two wave packets 
prepared at the same initial state and evolving forward in time under different 
Hamilton operators.

In general, time ordering makes the exact evaluation of $M(t)$ for a time
dependent Hamiltonian a daunting task. To circumvent this difficulty we
employ the semiclassical approximation, in which time ordering is trivially
accounted for by taking the time evolution of classical trajectories, as we 
detail in the sequel.

We consider the Hamiltonian $H$ defined as
\begin{equation}
H=H_{0}+V(\mathbf{q},t),
\end{equation}
where $H_{0}$ is a time independent Hamiltonian that displays chaotic
motion in the classical limit and $V({\bf q}, t)$ is the time-dependent
perturbation potential or the system interaction with a complex environment. 

The semiclassical propagator reads
\begin{eqnarray}
\langle\mathbf{q}^{\prime}|U(t)|\mathbf{q}\rangle &=& \left(  \frac{1}{2\pi\hbar
i}\right)  ^{d/2}\!\!\sum_{s(\mathbf{q}^{\prime},\mathbf{q},t)}C_{s}^{1/2}
\nonumber \\ & & \times 
\exp\left(  \frac{\mathrm{i}}{\hbar}S_{s}(\mathbf{q}^{\prime},\mathbf{q}
,t)-\frac{\mathrm{i}\pi}{2}\alpha_{s}\right),
\end{eqnarray}
where $s$ is a classical path that spends a time $t$ to travel from
$\mathbf{q}$ to $\mathbf{q}^{\prime}$, $S_{s}$ is the action (Hamilton
principal function), given by $S_{s}(\mathbf{q}^{\prime},\mathbf{q}
,t)=\int_{0}^{t}\mathrm{d}\tau L(\dot{\mathbf{q}}_{s}(\tau),\mathbf{q}
_{s}(\tau),\tau),$ $\alpha_{s}$ is the number of conjugate points along $s$,
and $C_{s}$ is the Jacobian of the phase-space transformation between
$\delta\mathbf{p}^{\prime}(0)$ and $\delta\mathbf{q}^{\prime}(t)$ -- a
density of classical paths.

It is only possible to proceed analytically if we restrict ourselves to the
regime of weak perturbations, in the sense that classical perturbation theory is
applicable.
More specifically, we approximate the action along a given trajectory $s$ by
\begin{equation}
S_{s}(t)\simeq S_{s}^{0}(t)+\int_{0}^{t}\mathrm{d}t^{\prime}\,V(\mathbf{q}%
_{s}(t^{\prime}),t^{\prime}), \label{action}%
\end{equation}
where $S_{s}^{0}(t)$ refers to the action corresponding to $s$ obtained 
from $H_{0}$ and $\mathbf{q}_{s}(t)$ gives the particle position along
the unperturbed trajectory $s$ as a function of time. 
For chaotic systems, this approximation is accurate up to a time 
$t_{\mathrm{cp}}$ proportional to the logarithm of the strength of $V$. 
In this sense, the perturbation is weak when $t_{\mathrm{cp}}$ becomes 
the largest time scale of the problem. This restriction does not preclude the 
perturbation to be 
quantum mechanically large, since the actions are measured in units of 
$\hbar$ \cite{CookSmooth}. It has been observed that the classical 
perturbation approximation works, in general, surprisingly well even for 
times longer than $t_{\mathrm{cp}}$. This has been related to the structural 
stability of the manifold of trajectories in phase space \cite{Cerruti02}: 
Even though individual trajectories are exponentially 
sensitive to perturbations, one can always find a \textquotedblleft 
replacement\textquotedblright\ trajectory in the manifold that joins the 
points of interest for a given time interval \cite{Vanicek03}.

Our calculation proceeds along the lines of Ref.~\onlinecite{Jalabert01},
which we now briefly sketch.
We assume that the wave packet $\langle\mathbf{r}|\psi_{0}\rangle$ is well 
localized, $\xi \gg\sigma\gg\lambda_{dB}$, where $\xi$ is a typical length 
of the perturbation (in Ref. \onlinecite{Jalabert01} 
the width of Gaussian impurities) and $\lambda_{dB}$ is the de Broglie wavelength of 
the particle. Neglecting terms with a rapidly oscillating phase, one arrives at 
the semiclassical expression for the Loschmidt echo,
\begin{eqnarray}
M(t)& \simeq & \left(  \frac{\sigma^{2}}{\pi\hbar^{2}}\right)  ^{d}\left\vert
\int\!d\mathbf{r}\sum_{s(\mathbf{r},\mathbf{r}_{0},t)}C_{s} 
\exp \left[
\frac{\mathrm{i}}{\hbar}\Delta S_{s}(t)\right] 
\right.
 \nonumber \\ & & \times
 \left.
  \exp\left[  -\frac{\sigma^{2}%
}{\hbar^{2}}\left(  \mathbf{p}_{s}-\mathbf{p}_{0}\right)  ^{2}\right]
\right\vert ^{2}, \label{MSC}
\end{eqnarray}
where $\Delta S_{s}$ is the action difference between trajectories evolved
with $H_{0}$ and $H$, and $\mathbf{p}%
=-\partial S_{s}/\partial\mathbf{r}|_{\mathbf{r}=\mathbf{r}_{0}}$. All
trajectories $s$ start at $\mathbf{r}_{0}$, the position where the Gaussian
wave packet $\langle\mathbf{r}|\psi_{0}\rangle$ is centered at. For short times 
$M(t)$ can only capture the local instabilities of the classical dynamics, thus,
it shows large fluctuations \cite{CookSmooth}. By
sampling over different initial values of $\mathbf{r}_{0}$ and $\mathbf{p}_{0}$,
or over an ensemble of perturbations, one obtains an average 
$\langle M(t)\rangle$ that puts in evidence the exponential decay \cite{CookLorentz03}. 
The Lyapunov and the FGR decay regimes are related to the different ways of pairing 
the path summations in the double sum of Eq.~(\ref{MSC}).

\subsection{Non diagonal contributions to $\langle M(t) \rangle$}

Let us first calculate the terms where the two trajectories lie far apart 
in phase space. Such contributions to $\langle M (t)\rangle$ are usually called 
non-diagonal (different trajectories), and read 
\begin{eqnarray}
\langle M^{nd}(t)\rangle & \simeq & \left(  \frac{\sigma^{2}}{\pi\hbar^{2}}\right)
^{d}\left\vert \int\mathrm{d}\mathbf{r}\sum_{s(\mathbf{r},\mathbf{r}_{0}%
,t)}C_{s} 
\left\langle \exp\left[  \frac{\mathrm{i}}{\hbar}\Delta
S_{s}(t)\right]  \right\rangle 
\right.
\nonumber \\ & & \times \left.
\exp\left[  -\frac{\sigma^{2}}{\hbar^{2}%
}\left(  \overline{\mathbf{p}}_{s}-\mathbf{p}_{0}\right)  ^{2}\right]
\right\vert ^{2},
\end{eqnarray}
where $\left\langle ...\right\rangle $ indicates that we average over the wave packet 
initial positions ${\bf r}_0$, as well as over an ensemble of perturbations.

We assume, as is customary for chaotic systems, that the actions for different paths 
are uncorrelated and Gaussian distributed \cite{Ozorio98,Vanicek03}. 
This leads to an enormous simplification, allowing us to write
\begin{equation}
\left\langle \exp\left[  \frac{\mathrm{i}}{\hbar}\Delta S_{s}(t)\right]
\right\rangle \simeq\exp\left[  -\frac{1}{2\hbar^{2}}\left\langle [\Delta
S_{s}(t)]^{2}\right\rangle \right]  .
\end{equation}
We remain with the task of evaluating the action variance
\begin{equation}
\left\langle \lbrack\Delta S_{s}(t)]^{2}\right\rangle =\int_{0}^{t}%
\mathrm{d}t^{\prime}\int_{0}^{t}\mathrm{d}t^{\prime\prime}%
\big\langle V(\mathbf{q}_{s}(t^{\prime}),t^{\prime})V(\mathbf{q}_{s}%
(t^{\prime\prime}),t^{\prime\prime})\big\rangle. \label{eq:actionvariance}%
\end{equation}
For that purpose we introduce an ensemble of perturbations $V$ to
model the general features of the environment. 
We replace the phase space average $\langle \cdots \rangle$ by the 
ensemble average $\overline{\cdots}$, the equivalence between averages being 
supported by the ergodicity of the system.
In order to keep our calculation as general as possible, we assume very little knowledge 
of the perturbation, requiring only that time and space correlations
are independent, viz.
\begin{equation}
\overline{V(\mathbf{q},t)V(\mathbf{q}^{\prime},t^{\prime})}=\overline{V^{2}%
}\,C_{S}(\left\vert \mathbf{q}-\mathbf{q}^{\prime}\right\vert )C_{T}%
(\left\vert t-t^{\prime}\right\vert ). 
\label{eq:defVstatistics}%
\end{equation}
The typical perturbation strength is $(\overline
{V^{2}})^{1/2}$, and $\tau_{V}=\hbar/(\overline{V^{2}})^{1/2}$ is its 
associated time scale. The dimensionless functions $C_{S}$ and $C_{T}$ 
quantify the spatial and time correlations of the potential $V({\bf q},t)$. 
We further require that $C_{S}$ or $C_{T}$ decay sufficiently fast, so that
\begin{equation}
\int_{0}^{\infty}\!\!\mathrm{d}r\,r^{d-1}C_{S}(r)<\infty\quad\mathrm{and}%
\quad\int_{0}^{\infty}\!\!dt\,C_{T}(t)<\infty .
\label{correlations}
\end{equation}
For chaotic systems this is a sensible assumption. 

To guide the discussion, let us introduce the correlation length $\xi_0$ and 
the correlation time $\tau_0$ that characterize $C_S$ and $C_T$ respectively. Since the 
average (\ref{eq:actionvariance}) is computed along the classical trajectories of the system, 
the asymptotic decay (\ref{correlations}) can be induced not only by the fluctuations of $V$, 
but also by the intrinsic chaotic dynamics of $H_0$. 
In general, $\xi_0$ and $\tau_0$ are given by the minimum between the natural scales of $V$ 
and $H_{0}$. For instance, when the perturbation is a static change in the mass tensor of a 
free particle bouncing off the walls of a billiard system, $\xi_0$ is solely given by the 
dynamics of $H_0$ and is equal to the mean free path between collisions 
\cite{CookLorentz, Wisniacki02,CookLorentz03}. Another example can be found in Refs.
\onlinecite{Jacquod01} and \onlinecite{Benenti02}, where the effective scale $\tau_0$ is given 
by the kicking period of the unperturbed Hamiltonian -- although the perturbation is a time 
independent change in 
the kicking strength. Hence, our results are valid not only for random
perturbations, but also for static and periodic ones: the chaoticity of the underlying 
Hamiltonian alone can enforce conditions (\ref{correlations}).

In the limit of $\tau_{0}\gg1/\lambda$ the perturbation is quasi-static and the results 
of Ref.\ \onlinecite{Jalabert01} hold without further change. 
We are interested in the regime where the typical times of the perturbation are comparable 
to those of the system, $\tau_{0}\lesssim1/\lambda.$

Replacing space averages by ensemble averages (\ref{eq:defVstatistics}), 
we write Eq.\ (\ref{eq:actionvariance}) as
\begin{eqnarray}
\left\langle \Delta S_{s}(t)^{2}\right\rangle & = & \overline{V^{2}}\int_{0}%
^{t}d\overline{t}\int_{-\infty}^{\infty}d\tau 
\nonumber \\ &  \times  &
C_{R}\Big(\left\vert
\mathbf{q}_{s}(\overline{t}-\frac{\tau}{2})-\mathbf{q}_{s}(\overline{t}+
\frac{\tau}{2})\right\vert \Big)\,C_{T}(\tau), \label{actionVariance}%
\end{eqnarray}
where we considered times $t$ much larger than $\tau_{0}$ and $\xi_0/v$, which
allows us to take the integral in $\tau$ from $-\infty$ to $+\infty$. 
Eq.~(\ref{actionVariance}) has two limiting regimes that are readily solved. 
In the first one, the spatial disorder has a much shorter scale than the 
temporal one: $\tau_{0}\gg\xi_0/v=\tau_\xi$. In this case the decay of 
$M^{nd}(t)$ is dominated by the same exponent as the one found in 
Ref. \onlinecite{Jalabert01}, 
\begin{align}
\left\langle \Delta S_{s}(t)^{2}\right\rangle  &  \simeq\overline{V^{2}}%
\int_{0}^{t}d\overline{t}\int_{-\infty}^{\infty}d\tau C_{S}\left[  \left\vert
\mathbf{q}_{s}(\overline{t}-\frac{\tau}{2})-\mathbf{q}_{s}(\overline{t}+
\frac{\tau}{2})\right\vert \right] \nonumber\\
&  =\frac{t}{\tilde{\tau}_{1}}\hbar^{2},
\end{align}
where $C_{T}(\tau)$ is assumed constant and $\tilde{\tau}$ is given by
a FGR calculation 
\begin{equation}
\frac{1}{\tilde{\tau}_{\mathrm{1}}}=\frac{\tau_{\xi}}{\tau_{V}^{2}}.
\end{equation}
When $\tau_{0}\ll\tau_{\xi},$ we deal the opposite
regime, and
\begin{equation}
\left\langle \Delta S_{s}(t)^{2}\right\rangle \simeq\overline{V^{2}}\int
_{0}^{t}d\overline{t}\int_{-\infty}^{\infty}d\tau C_{T}(\tau)=\frac
{t}{\tilde{\tau}_{2}}\hbar^{2},
\end{equation}
with
\begin{equation}
\frac{1}{\tilde{\tau}_{2}}=\frac{\tau_{0}}{\tau_{V}^{2}}.
\end{equation}

Thus, in these two limits and complementary situations, the FGR exponent changes from 
being governed by the spatial to the temporal correlations of $V({\bf q},t)$. 
The interesting ``correlation crossover regime" -- where neither the temporal nor the 
spatial correlation dominate -- will be discussed shortly for a particular form of 
$C_{S}$ and $C_{T}$.

\subsection{Diagonal contributions to $\langle M(t) \rangle$}

Let us first explicitly write (\ref{MSC}), namely 
\begin{eqnarray}
M(t)& \simeq& \left(  \frac{\sigma^{2}}{\pi\hbar^{2}}\right)  ^{d}\int
\!\mathrm{d}\mathbf{r}\int\!\mathrm{d}\mathbf{r}^{\prime}\!\!\sum_{
\genfrac{}{}{0pt}{}{{s(\mathbf{r},\mathbf{r}_{0},t)}}{{s^{\prime
}(\mathbf{r^{\prime}},\mathbf{r}_{0},t)}}
}C_{s}C_{s^{\prime}}
\nonumber \\ & & \times
\left\langle \exp\left[  \frac{i}{\hbar}\left(  \Delta
S_{s}(t)-\Delta S_{s^{\prime}}(t)\right)  \right]  \right\rangle \nonumber \\
& & \times \exp\left[
-\frac{\sigma^{2}}{\hbar^{2}}\left(  \left(  \overline{\mathbf{p}}
_{s}-\mathbf{p}_{0}\right)  ^{2}+\left(  \overline{\mathbf{p}}_{s}
-\mathbf{p}_{0}\right)  ^{2}\right)  \right]  , \label{eq-semiclassicM}
\end{eqnarray}
and analyze the case where the trajectories $s$ and $s^\prime$ remain 
close to each other. 
Now the action differences cannot be considered as uncorrelated, and 
we have to take into account the fluctuations in
\begin{eqnarray}
\left\langle \exp\left[  \frac{i}{\hbar}\left(  \Delta S_{s}(t)-\Delta
S_{s^{\prime}}(t)\right)  \right]  \right\rangle \simeq
\nonumber \\
\exp\left[  -\frac
{1}{2 \hbar^{2}}\left\langle \left[  \Delta S_{s}(t)-\Delta S_{s^{\prime}%
}(t)\right]  ^{2}\right\rangle \right]  . \label{phasefactor}%
\end{eqnarray}
In the same order of approximation of Eq.~(\ref{action}), we write
\begin{equation}
\Delta S_{s}(t)-\Delta S_{s^{\prime}}(t)=\int_{0}^{t}\mathrm{d}t^{\prime
}\text{ }\left[  V(\mathbf{q}_{s}(t^{\prime}),t^{\prime})-V(\mathbf{q}%
_{s^{\prime}}(t^{\prime}),t^{\prime})\right]  . \label{eq-ActionDiferences}%
\end{equation}
As the two trajectories remain close in coordinate space, we can expand 
$V({\bf q}_s(t), t)$ to first order around $s$ and obtain
\begin{equation}
\Delta S_{s}(t)-\Delta S_{s^{\prime}}(t)\simeq\int_{0}^{t}dt^{\prime}\text{
}\nabla V\left(  \mathbf{q}_{s}(t^{\prime}),t^{\prime}\right)  \cdot\left[
\mathbf{q}_{s}(t^{\prime})-\mathbf{q}_{s^{\prime}}(t^{\prime})\right].
\end{equation}
To calculate the action difference variance we turn our attention to the
force correlation function, namely 
\begin{equation}
C_{\nabla}(|\mathbf{q}-\mathbf{q}^{\prime}|,|t-t^{\prime}|)  \equiv \left\langle
\nabla V\left[  \mathbf{q},t\right]  \cdot\nabla V\left[  \mathbf{q}^{\prime
},t^{\prime}\right]  \right\rangle .
\end{equation}
As before, we introduce an ensemble of perturbations, and write
\begin{equation}
C_{\nabla}(|\mathbf{q}-\mathbf{q}^{\prime}|,|t-t^{\prime}|)  
=\overline{V^{2}}C_{T}(|t-t^{\prime}|)(\nabla_{\mathbf{q}}\cdot
\nabla_{\mathbf{q}^{\prime}})C_{S}(|\mathbf{q}-\mathbf{q}^{\prime}|),
\end{equation}
such that $(\nabla_{\mathbf{q}}\cdot\nabla_{\mathbf{q}^{\prime}}
)C_{S}(|\mathbf{q}-\mathbf{q}^{\prime}|)$ decays sufficiently fast, in the
sense defined by Eq.\ (\ref{correlations}).

As time evolves, the separation between the coordinates ${\bf q}_s(t)$
and ${\bf q}_{s^\prime}(t)$ grows as $e^{\lambda t}$, where $\lambda $ is
the largest Lyapunov exponent of $H_0$. As a result, after some algebra, 
Eq.~(\ref{phasefactor}) gives 
$\exp{[-A(\mathbf{r}-\mathbf{r}^{\prime})^{2}/\hbar^{2}]}$, with
\begin{equation}
A=\overline{V^{2}}\tau_{0}\frac{(1-e^{-2\lambda t})}{2\lambda} \label{AT}%
\end{equation}
when $C_{T}$ dominates the decay of $C_{\nabla}$, and
\begin{equation}
A=\overline{V^{2}}\frac{1-e^{-2\lambda t}}{2\lambda v}\int_{-\infty}^{\infty
}\mathrm{d}q\left[  \frac{1-d}{q}\frac{\partial C_{S}(q)}{\partial q}%
-\frac{\partial^{2}C_{S}(q)}{\partial q^{2}}\right]  , \label{AS}%
\end{equation}
when $C_{T}$ decays slowly. 

In summary, the main result of Ref. \onlinecite{Jalabert01} holds, namely
\begin{equation}
M(t)=\overline{A}\exp(-\lambda t)+B\exp(-t/\tilde{\tau}),
\end{equation}
where $\overline{A}=[m \sigma/(A^{1/2}t)]^{d}$, $\lambda$ is the classical
Lyapunov exponent of the system and $1/\tilde{\tau}$ is given by
Eq.\ (\ref{actionVariance}). The exponential decay of the LE is dominated by the
smallest between $1/\tilde{\tau}$ and $\lambda$, giving a crossover from
FGR to Lyapunov decay as the perturbation strength increases.

\subsection{Correlation crossover}

In the regime where $\tau_{0}\approx \tau_\xi$, one can only 
obtain further insight by assuming a specific form of the correlation functions.
Although it is a less general result, one can still encompass a broad class of
possible perturbations whose correlator decay in a particular way. We will
consider the case where both $C_{S}$ and $C_{T}$ have Gaussian shapes,
\begin{equation}
\overline{V(\mathbf{q},t)V(\mathbf{q}^{\prime},t^{\prime})}=
\frac{\overline{V^{2}}}{\pi}
\exp\left(-\frac{|\mathbf{q}-\mathbf{q}^{\prime}|^{2}}{\xi_0^{2}}\right) 
\exp\left(-\frac{|t-t^{\prime}|^{2}}{\tau_0^{2}}\right).
\end{equation}
Under the assumption that $t$ is large compared to $\tau_{0}$ and $\tau_\xi$, we
replace in Eq.~(\ref{actionVariance}) and Eq.~(\ref{phasefactor}), and obtain
the decay rate for the FGR regime
\begin{equation}
\frac{1}{\widetilde{\tau}} =\frac{\tau_{V}^{-2}}{\sqrt{\tau
_{0}^{-2}+\tau_{\xi}^{-2}}}\label{FGR-general},
\end{equation}
and the prefactor $A$ of the Lyapunov regime:
\begin{equation}
A=
\left( \frac{\hbar^2}{v^2 \lambda \widetilde{\tau}^3} \right)
\left( \frac{1-e^{-2\lambda t}}{\sqrt{\pi}} \right)
\tau_V^4 \left( \frac{d-1}{\tau_\xi^{4}} + \frac{d}{\tau_0^{2}\tau_\xi^{2}} \right).
\end{equation}
When the temporal or spatial correlation dominate, we recover the previous limit
\begin{equation}
\frac{1}{\widetilde{\tau}}\simeq \frac{\tau_{c}}{\tau_{V}^{2}}~~\mathrm{with}%
~~\tau_{c}=\min[\tau_{0}^{{}},\tau_{\xi}^{{}}] .\label{FGRtime-general}%
\end{equation}
As before, if the effective time scale $\tau_c$ becomes too short, the perturbation cancels 
itself out causing a very slow decay. This result is consistent with studies of time 
dependent errors in a quantum computer 
\cite{Pascazio}, where the dynamical decoupling to the environment was interpreted as a 
manifestation of the quantum Zeno effect \cite{QZE98, Facchi-Pascazio-QZspaces}.
Notice that when $\tau_c$ is dominated by the dynamics of $H_0$, the 
fluctuations become faster for chaotic systems with a larger $\lambda$
 \cite{CookLorentz03}.

\section{Conclusions}

We have extended the semiclassical theory of the Loschmidt echo to cope 
with time dependent perturbations. We expect our results to remain
valid in more complex or analytically difficult cases, suitable only for
numerical studies. 
Our treatment is sufficiently general as to describe the
situations where the perturbation is the random effect of an uncontrolled
environment on the system. The fluctuations we considered could arise either
from an explicit time dependence of the perturbation potential, or from the
ergodic nature of $H_{0}.$ In the last case, the underlying chaotic dynamics
mimics the randomness required for the decay of the
correlation functions. 
Thus, our results should also apply to periodic or very simple oscillating perturbations.

We showed that the Loschmidt echo Lyapunov regime is barely affected by the
time-dependence of the perturbation, except for prefactors: The decay is 
dominated by the system's intrinsic dynamics of stretching and folding.
In the FGR regime -- when the non-diagonal terms dominate -- the spatial and 
time scales of the perturbation compete with each other, and a simple behavior can 
be extracted when the relevant scales are far apart. 
In the intermediate regime, where the scales are comparable, using a 
simple (yet general) example we compute the decay rate of 
$M(t)$. 
The form of Eq. ($\ref{FGRtime-general}$) stresses how fast fluctuations 
lead to self-cancellation of the interaction with the environment. 
In the case of the LE, a vanishing FGR
exponent prevents the appearance of the perturbation independent Lyapunov regime.
Surprisingly, this happens not only for rapidly fluctuating perturbations, but also by 
increasing the Lyapunov exponent.
The slowing down of the FGR regime of decoherence -- induced by fast fluctuations -- was 
recently experimentally measured in NMR experiments \cite{Experimental-FGR}, 
where a connection to the quantum Zeno effect was observed.
It is interesting to recall that dynamical decoupling to the environment is what 
makes liquid NMR quantum computers possible (albeit small). The fast random
movements of the molecules in the liquid average out the more difficult to control
dipolar interactions present, e.g., in solids. 
Our work points to the importance of exploring dynamical alternatives to
suppress quantum decoherence \cite{Facchi-Pascazio-QZspaces}.

This work was initiated under a cooperation grant from\ Fundaci\'{o}n
Antorchas and Funda\c{c}\~{a}o Vitae. Further support from CNPq (Brazil) and
\ CONICET, ANPCyT and SeCyT-UNC (Argentina) is acknowledged.

\end{document}